\documentclass[
  nojss]{jss}

\usepackage{orcidlink,thumbpdf,lmodern}

\usepackage[utf8]{inputenc}

\author{
Tomasz Woźniak~\orcidlink{0000-0003-2212-2378}\\University of Melbourne
}
\title{Fast and Efficient Bayesian Analysis of Structural Vector
Autoregressions Using the \proglang{R} Package \pkg{bsvars}
\linebreak (Version 4.0)}

\Plainauthor{Tomasz Woźniak}
\Plaintitle{Fast and Efficient Bayesian Analysis of Structural Vector
Autoregressions Using the R Package bsvars}
\Shorttitle{\pkg{bsvars}: Bayesian Estimation of Structural Vector
Autoregressions}

\Abstract{
\noindent The \proglang{R} package \pkg{bsvars} provides a wide range of
tools for empirical macroeconomic and financial analyses using Bayesian
Structural Vector Autoregressions. It uses frontier econometric
techniques and \proglang{C++} code to ensure fast and efficient
estimation of these multivariate dynamic structural models, possibly
with many variables, complex identification strategies, and non-linear
characteristics. The models can be identified using adjustable exclusion
restrictions and heteroskedastic or non-normal shocks. They feature a
flexible three-level equation-specific local-global hierarchical prior
distribution for the estimated level of shrinkage for autoregressive and
structural parameters. Additionally, the package facilitates predictive
and structural analyses using impulse responses, forecast error variance
and historical decompositions, forecasting, statistical verification of
identification and hypotheses on autoregressive parameters, and analyses
of structural shocks, volatilities, and fitted values. These features
differentiate \pkg{bsvars} from existing \proglang{R} packages that
either focus on a specific structural model, do not consider
heteroskedastic shocks, or lack the implementation using compiled code.
}

\Plainkeywords{Bayesian inference, Structural VARs, Gibbs
sampler, exclusion restrictions, heteroskedasticity, non-normal
shocks, forecasting, structural analysis, R}

\Address{
    Tomasz Woźniak\\
    University of Melbourne\\
    Department of Economics\\
111 Barry Street\\
3053 Carlton, VIC, Australia\\
  E-mail: \email{tomasz.wozniak@unimelb.edu.au}\\
  URL: \url{https://bsvars.org}\\~\\
  }

\usepackage[utf8]{inputenc} \usepackage{amsmath} \usepackage{amssymb} \usepackage{natbib} \usepackage{booktabs} \usepackage{multirow} \usepackage{tikz}

\usetikzlibrary{shapes,positioning,arrows.meta} \tolerance=1
\emergencystretch=\maxdimen \hyphenpenalty=10000 \hbadness=10000

\begin{document}

\section{Introduction}

Since the publication of the seminal paper by
\cite{sims1980macroeconomics} Structural Vector Autoregressions (SVARs)
have become benchmark models for empirical macroeconomic analyses.
Subsequently, they have found numerous applications in other fields and
are now indispensable in everyday work at central banks, treasury
departments and other economic governance institutions, as well as in
finance, insurance, banking, and economic consulting.

The great popularity of these multivariate dynamic structural models was
gained because they incorporate the reduced and structural forms into a
unified framework. On the one hand, they capture the essential
properties of macroeconomic and financial time series such as
persistence, dynamic effects, system modelling, and potentially
time-varying conditional variances. On the other hand, they control for
the structure of an economy, system, or market through the
contemporaneous effects and, thus, they identify contemporaneously and
temporarily uncorrelated shocks that can be interpreted structurally.
All these features make it possible to estimate the dynamic causal
effects of the shocks on the measurements of interest reliably. These
effects are interpreted as the propagation of the well-isolated and
unanticipated cause -- a structural shock -- in the considered system of
variables throughout the predictable future.

This flexibility comes at a cost of dealing with local identification of
the model, sharply growing dimension of the parameter space with the
increasing number of variables, and the estimation of latent variables.
Bayesian inference provides original solutions to each of these
challenges often deciding on the feasibility of the analyses with a
demanded model including many variables, conditional heteroskedasticity,
and sophisticated identification of the structural shocks. In this
context, Markov Chain Monte Carlo methods grant certainty of reliable
estimation thanks to simple convergence diagnostics but they might incur
substantial computational cost.

The paper at hand and the corresponding package \pkg{bsvars} by
\cite{bsvars} for \proglang{R} \citep{Rcore} provide tools for empirical
macroeconomic and financial analyses using Bayesian SVARs. It addresses
the considered challenges by choosing a convenient model formulation,
applying frontier econometric and numerical techniques, and relying on
compiled code written using \proglang{C++} to ensure fast and efficient
estimation. Additionally, it offers a great flexibility in choosing the
model specification and identification pattern, modifying the prior
assumptions, and accessing interpretable tabulated or plotted outputs.
Therefore, the package makes it possible to benefit from the best of the
two facilities: the convenience of data analysis using \proglang{R} and
the computational speed using pre-compiled code written in
\proglang{C++}.

More specifically, the package uses the SVAR models featuring a standard
reduced form VAR equation following \cite{Banbura2010}
\citep[see also][]{Wozniak2016} and a structural equation linking the
reduced form error term to the structural shocks via the structural
matrix as in \cite{LSUW2024} and \cite{chankoopyu2024}. The normal prior
distribution for the autoregressive parameters implements the
interpretability of the Minnesota prior by \cite{Doan1984} by centring
it around the mean that reflects unit-root nonstationarity or
stationarity of the variables with an adjustable level of shrinkage
depending on the equation and exhibiting exponential decay with the
increasing autoregressive lag order. The prior distribution for the
structural matrix is generalised-normal by \cite{WaggonerZha2003} which
preserves the shape of the likelihood function. Both of these priors are
combined with the flexible three-level equation-specific local-global
hierarchical prior distribution for the estimated level of shrinkage as
in \cite{LSUW2024} improving the model fit. The estimation of the level
of shrinkage was shown to substantially improve forecasting performance
of VAR models by \cite{Giannone2015}. Additionally, these specification
choices lead to an efficient equation-by-equation Gibbs sampler for the
posterior distribution of the autoregressive and structural parameters
proposed by \cite{chankoopyu2024} and \cite{WaggonerZha2003}
respectively.

\begin{figure}
\centering
\includegraphics[width=2in,height=2.314in,alt={The hexagonal package logo features an impulse response that can be fully reproduced using the  package following a script available at: }]{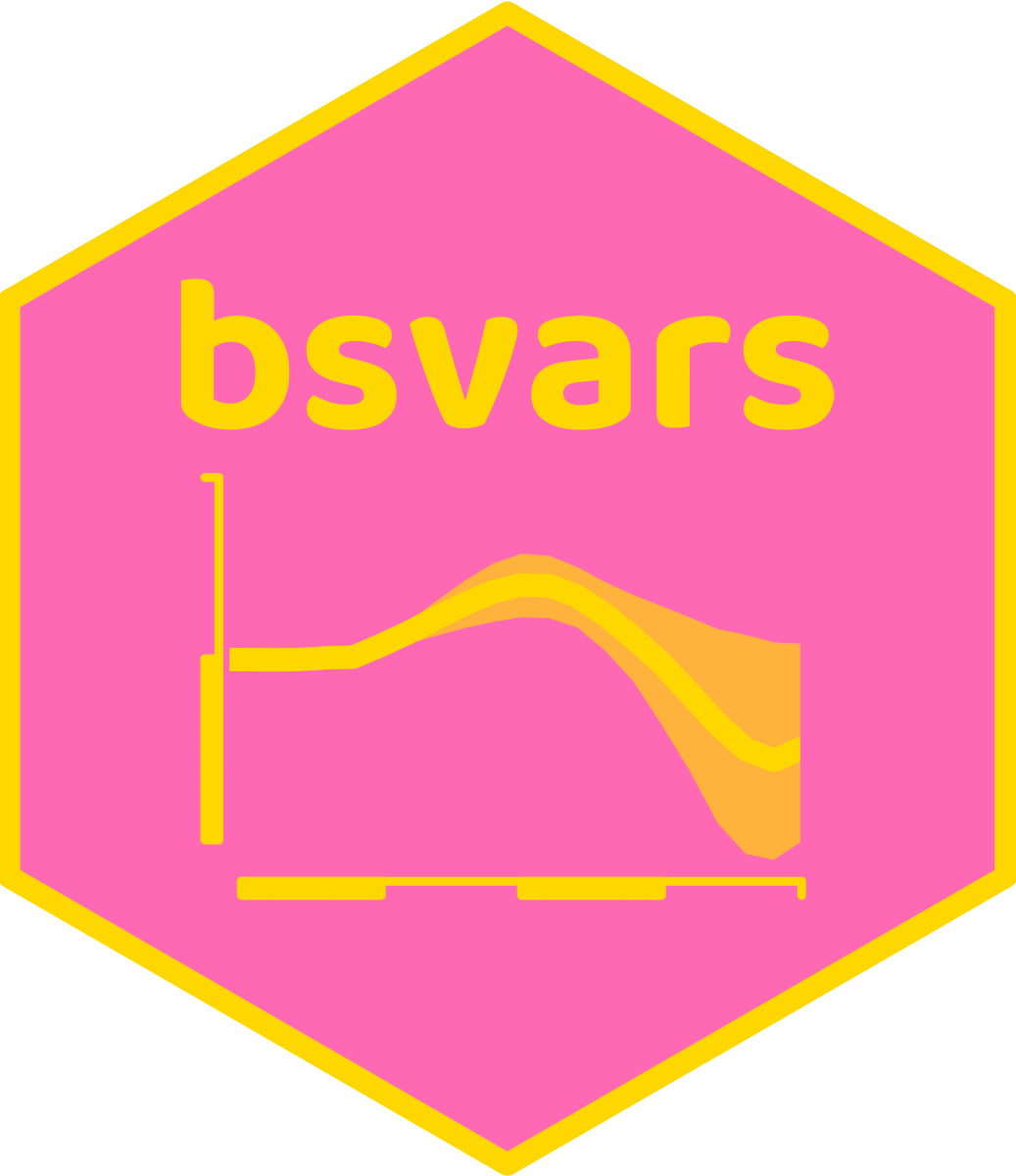}
\caption{The hexagonal package logo features an impulse response that
can be fully reproduced using the \pkg{bsvars} package following a
script available at:
\href{https://github.com/bsvars/hex/blob/main/bsvars/bsvars.R}{github.com/bsvars/hex/blob/main/bsvars/bsvars.R}}
\end{figure}

All the structural models in the \pkg{bsvars} package can be identified
using highly adaptable exclusion restrictions for the structural and
autoregressive parameters proposed by \cite{WaggonerZha2003} and use the
structural matrix sign normalisation by \cite{WaggonerZha2003norm}.
However, one might also choose to include additional sources of
identification through heteroskedasticity following the ideas by
\cite{Rigobon03} or by non-normal shocks as proposed by
\cite{Lanne2010}. Therefore, the package offers a range of models for
structural shocks conditional variance including a homoskedastic model
with time-invariant variances. The list of heteroskedastic
specifications is opened by Stochastic Volatility (SV) model in two
versions: non-centred as in \cite{LSUW2024} and centred used by
\cite{chankoopyu2024}, and is followed by the Markov-switching
heteroskedasticity (MSH) model proposed by \cite{SW2026} and
\cite{brunnermeier2021}. The implementations of the MSH models are
available in two versions: with stationary Markov process and its sparse
version that facilitates the estimation of the number of states and
non-parametric interpretations. Similarly, models with non-normal shocks
are implemented in three versions including Student-t distributed shocks
proposed by \cite{lanne2017}, and those following the normal mixture
(MIX) model: with a finite number of components as in
\cite{FruhwirthSchnatter2006} and in its sparse representation serving
as an approximation of a non-parametric infinite mixture inspired by
\cite{malsiner2016model}.

The \proglang{R} package \pkg{bsvars} implements a wide range of tools
for structural and predictive analyses. The former encompasses the
methods comprehensively revised by
\citep[][Chapter 4: SVAR Tools]{KL2017} and include the impulse response
functions, forecast error variance decompositions, historical
decompositions, as well as the basic analysis of fitted values,
structural shocks, conditional standard deviations and regime
probabilities for MSH and MIX models. The predictive analysis includes
Bayesian forecasting implemented through an algorithm sampling from the
predictive density of the unknown future values of the dependent
variables, and conditional forecasting of a number of variables given
other variables' projections. Methods \texttt{summary()} and
\texttt{plot()} support the user in the interpretations and
visualisation of such analyses.

A distinguishing feature of the package are the Bayesian model
diagnostic tools for the verification of identification and hypotheses
on autoregressive parameters using posterior odds ratios. The posterior
odds for hypotheses expressed as sharp restrictions on parameters are
computed using Savage-Dickey Density Ratios (SDDRs) by
\cite{Verdinelli1995}. The SDDRs representing Bayes factors are
reliable, precisely estimated, and straightforward to compute once the
posterior sample is available. As shown by \cite{LW2017} and
\cite{LSUW2024}, SDDRs are a pragmatic tool for the Bayesian assessment
of a null hypothesis of homoskedasticity in heteroskedastic models
allowing to verify partial or global identification through
heteroskedasticity. A similar argument applies for the verification of
identification in non-normal models. Additionally, the package's general
implementation of SDDRs for the autoregressive parameters facilitates
verification of restrictions on any conditional mean parameters of the
model, such as e.g., hypotheses of exogeneity or Granger non-causality.

Finally, the package \pkg{bsvars} is highly integrated in terms of
workflows, objects, and code compatibility with \proglang{R} packages
\pkg{bsvarSIGNs} by \cite{bsvarSIGNs,WangWozniak2025} focusing on SVAR
models identified using narrative, sign, and zero restrictions, as well
as with \pkg{bvars} by \cite{bvars} and \pkg{bpvars} by \cite{bpvars}
and \cite{smw2026} for VAR and panel VAR analyses, respectively. These
packages usage compatibility and its complementarity in terms of the
implemented methods constitutes an additional appeal of the \pkg{bsvars}
package.

Multivariate dynamic modelling, both Bayesian and frequentist, has found
some traction in \proglang{R} in the recent years, which resulted in
many new packages available on the CRAN repository. Two packages,
\pkg{MTS} by \cite{MTS} and \pkg{vars} by \cite{vars}, cover a wide
range of benchmark models for multivariate time series analysis in
economics in finance. Other packages provide functionality for
reduced-form models estimation and forecasting implementing
regularisation, such as the \pkg{BigVAR} package by \cite{BigVAR},
\pkg{bigtime} by \cite{bigtime}, and \pkg{VARshrink} by
\cite{VARshrink}, or Bayesian shrinkage, such as the \pkg{bvartools}
package by \cite{bvartools}, \pkg{bayesianVARs} by \cite{bayesianVARs},
or \pkg{BGVAR} by \cite{BGVAR}. All of these specification are proven to
be highly beneficial for forecasting in particular contexts.

Notable implementations of structural models include packages focusing
on specific models important from the point of view of historical
developments in the field, namely, the package \pkg{BVAR} by \cite{BVAR}
providing tools for the estimation and analysis proposed by
\cite{Giannone2015}, package \pkg{bvarsv} by \cite{bvarsv} focusing on
the heteroskedastic VAR proposed by \cite{Primiceri2005}, and package
\pkg{FAVAR} by \cite{FAVAR} implementing the factor-augmented model by
\cite{Bernanke2005}. Other two packages that have been archived and are
no longer available on CRAN are the package \pkg{MSBVAR} by
\cite{MSBVAR} focusing on the Markov switching model by \cite{Sims2006}
and package \pkg{VARsignR} by \cite{VARsignR} provided a treatment of
Bayesian SVARs identified via sign restrictions by
\cite{uhlig2005effects}, \cite{rubio2010structural}, and
\cite{fry2011sign}. Some other packages focus on the implementation of
research code for families of models focused around a theme, such as the
aforementioned package \pkg{bsvarSIGNs} for sign-restricted SVARs,
package \pkg{gmvarkit} by \cite{gmvarkit} implementing frequentist SVARs
with non-Gaussian identification by \cite{KALLIOVIRTA2016485} and
\cite{Virolainen}, or package \pkg{sstvars} by \cite{sstvars} focusing
on smooth-transition non-linearity in the structural models by
\cite{virolainen2024}, \cite{ANDERSON19981}, and \cite{LUTKEPOHL2017}.
Importantly, there exists a whole universe of \proglang{MATLAB}
libraries for structural macroeconomics analyses. However, the code is
available mostly from authors' websites and without structured
documentation. This family of libraries is not surveyed here but we
single out the \pkg{BEAR} toolkit by \cite{BEAR} and \pkg{Dynare} by
\cite{Dynare} providing comprehensive set of methods and extensive
documentation.

However, the most relevant package to compare \pkg{bsvars} to is the
\pkg{svars} package by \cite{svars} focusing on frequentist inference
for SVAR models identified via exclusion restrictions,
heteroskedasticity, and non-normal shocks and implementing a range of
models that are feasible to estimate using the maximum likelihood
method. The similarity to the functionality of package \pkg{bsvars}
include the selection of models, such as the MSH and with non-normal
residuals, as well as the selection of tools for structural analyses
including impulse responses, historical and forecast error variance
decompositions. However, the package \pkg{svars} implements maximum
likelihood and bootstrap procedures for the analysis of the model
parameters and offers some specification testing procedures.

In this context, the \pkg{bsvars} package implements a range of novel
solutions and models for Bayesian analysis. One differentiating example
is the implementation of the SVAR models with SV that is not covered by
the package \pkg{svars}. This model is particularly important in the
context of recent developments clearly indicating that SV is the single
extension of VARs leading to marginally largest improvements in the
model fit and forecasting performance as shown e.g.~by \cite{Clark2015},
\cite{Chan2018}, \cite{carriero_large_2019}, \cite{chan2020large}, and
\cite{bertsche2022identification}. Another such example are sparse MSH
and MIX models based on hierarchical prior structures deciding on their
frequentist implementation infeasibility. Additionally, the package
\pkg{bsvars} provides unique Bayesian statistical procedures for the
verification of partial identification through heteroskedasticity and
non-normality using method \texttt{verify\_identification()}. The
verification of identification for such flexible models as those
considered in the package is not feasible in the frequentist framework.
Finally, the package benefits from the advantage of Bayesian approach
that facilitates the estimation for models with potentially many
variables, autoregressive lags inflating the dimension of parameters
space, Markov-switching regimes or normal mixture components, all of
which are the factors constraining the feasibility of maximum likelihood
approaches.

\section{Bayesian Analysis of Structural VARs}\label{sec:bayes}

This section scrutinises the modelling framework used in the package
focusing on the specification of the models, prior distributions,
hypotheses verification tools, and estimation. The reader is referred to
\citep[][Chapter 4: SVAR Tools]{KL2017} for the exposition of the
standard tools for the analysis of SVAR models, such as the impulse
responses, forecast error variance and historical decompositions, as
their implementation in the package closely follows this resource.

\subsection{Structural VARs}\label{ssec:svars}

All of the models in the package \pkg{bsvars} share the reduced and
structural form equations, as well as the hierarchical prior
distributions for these parameters following \cite{LSUW2024}. The
reduced form equation is the VAR equation with \(p\) lags specified for
an \(N\)-vector \(\mathbf{y}_t\) collecting observations on \(N\)
variables at time \(t\): \begin{align}
\mathbf{y}_t &= \mathbf{A}_1 \mathbf{y}_{t-1} + \dots + \mathbf{A}_p \mathbf{y}_{t-p} + \mathbf{A}_d \mathbf{d}_t +  \boldsymbol{\varepsilon}_t, \label{eq:var}
\end{align} where \(\mathbf{A}_i\) are \(N\times N\) matrices of
autoregressive slope parameters, \(\mathbf{d}_t\) is a \(D\)-vector of
deterministic terms, always including a constant term, and possibly
dummy and exogenous variables, \(\mathbf{A}_d\) is an \(N\times D\)
matrix of the corresponding parameters, and
\(\boldsymbol{\varepsilon}_t\) collects the \(N\) reduced form error
terms. Collect all the autoregressive matrices and the slope terms in an
\(N\times (Np+D)\) matrix
\(\mathbf{A} = \begin{bmatrix}\mathbf{A}_1& \dots & \mathbf{A}_p & \mathbf{A}_d\end{bmatrix}\)
and the explanatory variables in a \((Np+D)\)-vector
\(\mathbf{x}_t = \begin{bmatrix}\mathbf{y}_{t-1}' & \dots & \mathbf{y}_{t-p}' & \mathbf{d}_{t}' \end{bmatrix}'\).
Then equation \eqref{eq:var} can be written in the matrix form as
\begin{align}
\mathbf{y}_t &= \mathbf{A}\mathbf{x}_t + \boldsymbol{\varepsilon}_t. \label{eq:rf}
\end{align} Each of the rows of the matrix \(\mathbf{A}\), denoted by
\([\mathbf{A}]_{n\cdot}\), may feature exclusion restrictions
representing, for instance, exogeneity, small-open economy assumptions,
or no Granger causality hypothesis. The restrictions are imposed
following the approach by \cite{WaggonerZha2003} who decompose the
row-specific parameters into: \begin{align}
[\mathbf{A}]_{n\cdot} = \mathbf{a}_n\mathbf{V}_{A.n}, \label{eq:restrictionsA}
\end{align} where \(\mathbf{a}_n\) is a \(1\times r_{A.n}\) vector
collecting the elements to be estimated and the \(r_{A.n}\times N\)
matrix \(\mathbf{V}_{A.n}\) including zeros and ones placing the
estimated elements in the demanded elements of
\([\mathbf{A}]_{n\cdot}\). The unrestricted elements follow a
multivariate conditional normal prior distribution, given the
equation-specific shrinkage hyper-parameter \(\gamma_{A.n}\), with the
mean vector \(\mathbf{V}_{A.n}\underline{\mathbf{m}}_{n.A}\) and the
covariance
\(\gamma_{A.n}\mathbf{V}_{A.n}\underline{\Omega}_A\mathbf{V}_{A.n}'\),
denoted by: \begin{align}
\mathbf{a}_n'\mid\gamma_{A.n} \sim\mathcal{N}_{r_{A.n}}\left( \mathbf{V}_{A.n}\underline{\mathbf{m}}_{n.A}, \gamma_{A.n}\mathbf{V}_{A.n}\underline{\boldsymbol\Omega}_A\mathbf{V}_{A.n}' \right),\label{eq:priorA}
\end{align} where \(\underline{\mathbf{m}}_{n.A}\) is specified in-line
with the Minnesota prior by \cite{Doan1984} as a vector of zeros if all
of the variables are stationary, or containing value 1 in its
\(n\textsuperscript{th}\) element if the \(n\textsuperscript{th}\)
variable is unit-root nonstationary. By default,
\(\underline{\boldsymbol{\Omega}}_A\) is a diagonal matrix with vector
\(\begin{bmatrix}\mathbf{p}^{-2\prime}\otimes\boldsymbol{\imath}_N' & 100\boldsymbol{\imath}_D'\end{bmatrix}'\)
on the main diagonal, where \(\mathbf{p}\) is a vector containing a
sequence of integers from 1 to \(p\) and \(\boldsymbol\imath_N\) is an
\(N\)-vector of ones. \(\mathbf{V}_{A.n}\),
\(\underline{\mathbf{m}}_{n.A}\) and
\(\underline{\boldsymbol{\Omega}}_A\) can be modified by the user. This
specification includes the shrinkage level exponentially decaying with
the increasing lag order, relatively large prior variances for the
deterministic term parameters, and the flexibility of the hierarchical
prior that leads to the estimation of the level of shrinkage as proposed
by \cite{Giannone2015}. The latter feature is facilitated by assuming a
3-level local-global hierarchical prior on the equation-specific reduced
form parameters shrinkage given by \begin{align}
\gamma_{A.n} | s_{A.n}  &\sim\mathcal{IG}2\left(s_{A.n}, \underline{\nu}_A\right),\\
s_{A.n} | s_{A} &\sim\mathcal{G}\left(s_{A}, \underline{a}_A\right),\\
s_{A} &\sim\mathcal{IG}2\left(\underline{s}_{s_A}, \underline{\nu}_{s_A}\right),
\end{align} where \(\mathcal{G}\) and \(\mathcal{IG}2\) are gamma and
inverted gamma 2 distributions \citep[see][Appendix A]{Bauwens1999},
hyper-parameters \(\gamma_{A.n}\), \(s_{A.n}\), and \(s_{A}\) are
estimated, and \(\underline{\nu}_A\), \(\underline{a}_A\),
\(\underline{s}_{s_A}\), and \(\underline{\nu}_{s_A}\) are all set by
default to value 10 to assure appropriate level of shrinkage towards the
prior mean. The values of the hyper-parameters that are underlined in
our notation can be modified by the user.

The structural form equation determines the linear relationship between
the reduced-form innovations \(\boldsymbol{\varepsilon}_t\) and the
structural shocks \(\mathbf{u}_t\) using the \(N\times N\) structural
matrix \(\mathbf{B}_0\): \begin{align}
\mathbf{B}_0\boldsymbol{\varepsilon}_t = \mathbf{u}_t.\label{eq:sf}
\end{align} The structural matrix specifies the contemporaneous
relationship between the variables in the system and determines the
identification of the structural shocks from vector \(\mathbf{u}_t\).
Its appropriate construction may grant specific interpretation to one or
many of the shocks. The package \pkg{bsvars} facilitates the
identification of the structural matrix and the shocks via exclusion
restrictions \citep[see][Chapter 8]{KL2017} and/or through
heteroskedasticity or non-normal shocks \citep[][Chapter 14]{KL2017}.
The zero restrictions are imposed on the structural matrix row-by-row
following the framework proposed by \cite{WaggonerZha2003} via the
following decomposition of the \(n\textsuperscript{th}\) row of the
structural matrix, denoted by \([\mathbf{B}_0]_{n\cdot}\): \begin{align}
[\mathbf{B}_0]_{n\cdot} = \mathbf{b}_n\mathbf{V}_{B.n}, \label{eq:restrictions}
\end{align} where \(\mathbf{b}_n\) is a \(1\times r_{B.n}\) vector
collecting the elements to be estimated and the \(r_{B.n}\times N\)
matrix \(\mathbf{V}_{B.n}\) including zeros and ones placing the
estimated elements in the demanded elements of
\([\mathbf{B}_0]_{n\cdot}\).

The structural matrix \(\mathbf{B}_0\) follows a conditional
generalised-normal prior distribution by \cite{WaggonerZha2003} recently
revised by \cite{Arias2018} that is proportional to: \begin{align}
\mathbf{B}_0\mid\gamma_{B.1},\dots,\gamma_{B.N} \sim |\det(\mathbf{B}_0)|^{\underline{\nu}_B - N} \exp\left\{-\frac{1}{2} \sum_{n=1}^{N} \gamma_{B.n}^{-1} \mathbf{b}_n\underline{\boldsymbol\Omega}_{B.n}^{-1}\mathbf{b}_n' \right\},
\end{align} where \(\underline{\boldsymbol\Omega}_{B.n}\) is an
\(r_{B.n}\times r_{B.n}\) scale matrix set to the identity matrix by
default, \(\underline{\nu}_B \geq N\) is a shape parameter, and
\(\gamma_{B.n}\) is an equation-specific structural parameter shrinkage.
The shape parameter \(\underline{\nu}_B\) set to \(N\) by default makes
this prior a conditional, zero-mean \(r_{B.n}\)-variate normal prior
distribution for \(\mathbf{b}_n\) with the diagonal covariance and the
diagonal element \(\gamma_{B.n}\). The shape parameter can be modified
by the user though.

This prior specification is complemented by a 3-level local-global
hierarchical prior on the equation-specific structural parameters
shrinkage given by \begin{align}
\gamma_{B.n} | s_{B.n}  &\sim\mathcal{IG}2\left(s_{B.n}, \underline{\nu}_b\right),\\
s_{B.n} | s_{B} &\sim\mathcal{G}\left(s_{B}, \underline{a}_B\right),\\
s_{B} &\sim\mathcal{IG}2\left(\underline{s}_{s_B}, \underline{\nu}_{s_B}\right),
\end{align} where hyper-parameters \(\gamma_{B.n}\), \(s_{B.n}\), and
\(s_{B}\) are estimated and \(\underline{\nu}_b\), \(\underline{a}_B\),
\(\underline{s}_{s_B}\), and \(\underline{\nu}_{s_B}\) are fixed to
values 10, 10, 1, and 100 respectively to assure a flexible dispersed
distribution \emph{a priori} but they can be modified by the user.

Finally, all of the models share the zero-mean conditional normality of
the structural shocks given the past observations with the diagonal
covariance matrix containing the \(N\)-vector of structural shock
variances, \(\boldsymbol{\sigma}_t^2\), on the main diagonal:
\begin{align}
\mathbf{u}_t\mid\mathbf{x}_t, \boldsymbol{\sigma}_t^2 \sim\mathcal{N}_{N}\left( \mathbf{0}_N, \text{diag}\left(\boldsymbol{\sigma}_t^2\right) \right).\label{eq:ss}
\end{align} The diagonal covariance matrix, together with the joint
normality, implies contemporaneous independence of the structural shocks
which is the essential feature allowing for the estimation of dynamic
effects to a well-isolated cause that is not influenced by other factors
in the SVAR models.

The model parts described in the current section are common to all the
models considered in the package \pkg{bsvars}. Note that identification
of the structural matrix here must be assured by the exclusion
restrictions only in the homoskedastic model. Heteroskedastic and
non-normal specifications might not require the exclusion restrictions
to identify the structural matrix. Still, such restrictions might occur
beneficial from the point of view of the model fit sharpening shock
identification. Such alternative model specifications are distinguished
by the characterisation of the conditional variances collected in vector
\(\boldsymbol{\sigma}_t^2\).

\subsection{Models for Conditional Variances}\label{ssec:variances}

The \pkg{bsvars} package offers a selection of alternative
specifications for structural shocks conditional variance process.

\subsubsection{Homoskedastic model}

The first such specification is a homoskedastic model for which the
conditional variance of every shock is equal to \begin{align}
\sigma_{n.t}^2 = 1 \label{eq:homosk}
\end{align} for all \(t\). This setup results in a simple SVAR model
that is quick to estimate. Note that in this model, the conditional
covariance of the data vector, \(\mathbf{y}_t\) is equal to
\((\mathbf{B}_0'\mathbf{B}_0)^{-1}\). The point at which this model is
standardized as in equation \eqref{eq:homosk} is the benchmark value for
the all other heteroskedastic models whose conditional variances hover
around value 1.

\subsubsection{Stochastic Volatility}

The heteroskedastic model with Stochastic Volatility is implemented in
two versions: non-centred by \cite{LSUW2024} and centred
\cite{chankoopyu2024}. In these models, the conditional variances for
each of the shocks are given by \begin{align}
\text{non-centred:}&&\sigma_{n.t}^2 &= \exp\left\{\omega_n h_{n.t}\right\}\\
\text{centred:}&&\sigma_{n.t}^2 &= \exp\left\{\tilde{h}_{n.t}\right\},
\end{align} where the parameter \(\omega_n\) denotes the plus-minus
square root of the conditional variance for the log-conditional variance
process and will be referred to as the
\emph{volatility of the log-volatility}, whereas \(h_{n.t}\) and
\(\tilde{h}_{n.t}\) are the log-volatility processes following
autoregressive equations \begin{align}
\text{non-centred:}&&h_{n.t} &= \rho_n h_{n.t-1} + v_{n.t}, && v_{n.t}\sim\mathcal{N}\left(0,1\right)\\
\text{centred:}&&\tilde{h}_{n.t} &= \rho_n \tilde{h}_{n.t-1} + \tilde{v}_{n.t}, && \tilde{v}_{n.t}\sim\mathcal{N}\left(0,\sigma_v^2\right)
\end{align} with the initial values \(h_{n.0} = \tilde{h}_{n.0} = 0\),
where \(\rho_n\) is the autoregressive parameter, \(v_{n.t}\) and
\(\tilde{v}_{n.t}\) are the SV normal innovations, and \(\sigma_v^2\) is
the conditional variance of \(\tilde{h}_{n.t}\) in the centred
parameterisation.

The priors for these models include a uniform distribution for the
autoregressive parameter \begin{align}
\rho_n \sim\mathcal{U}(-1,1),
\end{align} which assures the stationarity of the log-volatility process
and sets its unconditional expected value to
\(E[h_{n.t}] = E[\tilde{h}_{n.t}] =0\). The prior distribution for the
volatility of the log-volatility parameter in the non-centred and the
conditional variance of the log-volatility in the centred
parameterisation follow a multi-level hierarchical structures given by:
\begin{align}
\text{non-centred:}&&\omega_n\mid \sigma_{\omega.n}^2 &\sim\mathcal{N}\left(0, \sigma_{\omega.n}^2\right),\qquad
\sigma_{\omega.n}^2 \mid s_\sigma\sim\mathcal{G}(s_\sigma, \underline{a}_\sigma), \qquad
s_\sigma \sim\mathcal{IG}2(\underline{s}, \underline{\nu})\\
\text{centred:}&&\sigma_v^2 \mid s_v &\sim\mathcal{IG}2(s_v, \underline{a}_v), \qquad
s_v \mid s_\sigma\sim\mathcal{G}(s_\sigma, \underline{a}_\sigma), \qquad
s_\sigma \sim\mathcal{IG}2(\underline{s}, \underline{\nu})
\end{align} where parameters \(\sigma_{\omega.n}^2\) and \(s_v\) follow
gamma distribution with expected value equal to
\(s_\sigma\underline{a}_\sigma\). In this hierarchical structure the
hyper-parameters \(\omega_n\), \(\sigma_{\omega.n}^2\), \(\sigma_v^2\),
\(s_v\), and \(s_\sigma\) are estimated, while \(\underline{a}_v\),
\(\underline{a}_\sigma\), \(\underline{\nu}\), and \(\underline{s}\) are
set to values 1, 1, 1, and 0.1, respectively, by default and can be
modified by the user.

\begin{CodeChunk}
\begin{figure}

{\centering \includegraphics{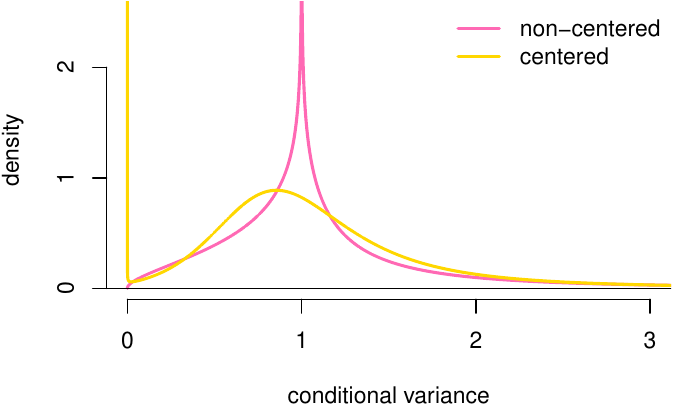} 

}

\caption[Marginal prior density of a structural shock conditional variance in the two SV models]{Marginal prior density of a structural shock conditional variance in the two SV models. Source: Lütkepohl et al. (2024).}\label{fig:plot_cv_prior}
\end{figure}
\end{CodeChunk}

Both of the SV models share a number of features, such as the
flexibility of hierarchical priors with estimated hyper-parameters
granting better fit to data and improved forecasting performance, and
they facilitate the identification of the structural matrix
\(\mathbf{B}_0\) and shocks through heteroskedasticity following the
original ideas by \cite{bertsche2022identification} and
\cite{lewis_identifying_2021}. However, they differ with respect to the
features analysed by \cite{LSUW2024} that are presented in Figure
\ref{fig:plot_cv_prior} plotting the marginal prior distribution of the
conditional variances, \(\sigma_{n.t}^2\), in the two SV models. These
densities are implied by the SV model specifications presented in this
section. The centred parameterisation concentrate the prior probability
mass around point 1 only mildly and goes to infinity when the
conditional variance goes to zero. The non-centred parameterisation, on
the other hand, concentrates the prior probability mass around point 1
more strongly, goes to zero when the conditional variance goes to zero,
and features fat right tail. Finally, in the non-centred
parameterisation homoskedasticity can be verified by checking the
restriction \(\omega_n = 0\) \citep[see][]{LSUW2024}.

\subsubsection{Markov-switching heteroskdasticity}

In another heteroskedastic model, the MSH one as proposed by
\cite{SW2026}, the time-variation of the conditional variances is
determined by a discrete-valued Markov process \(s_t\) with \(M\)
regimes: \begin{align}
\sigma_{n.t}^2 = \sigma_{n.s_t}^2.
\end{align} All of the variances switch their values at once according
to the latent Markov process takes the values
\(s_t = m\in\{1,\dots,M\}\). The properties of the Markov process itself
are determined by the transition matrix \(\mathbf{P}\) whose
\([\mathbf{P}]_{i.j}\) element denotes the transition probability from
regime \(i\) to regime \(j\) over the next period. The process' initial
probabilities are estimated and denoted by the \(M\)-vector
\(\boldsymbol{\pi}_0\). In this model used by \cite{brunnermeier2021},
the variances in the \(n\textsuperscript{th}\) equation sum to \(M\),
each of them has the prior expected value equal to 1, and their regimes
are given equal prior probabilities of occurrence equal to \(M^{-1}\).
Therefore, the prior for the conditional variances is the \(M\)-variate
Dirichlet distribution: \begin{align}
M^{-1}\left(\sigma_{n.1}^2, \dots, \sigma_{n.M}^2\right) \sim\mathcal{D}irichlet_M(\underline{e}_\sigma, \dots, \underline{e}_\sigma), \label{eq:sigmaMSprior}
\end{align} where the hyper-parameter \(\underline{e}_\sigma = 1\) is
fixed. Each of the rows of the transition matrix as well as the initial
state probabilities follow the Dirichlet distribution as well:
\begin{align}
[\mathbf{P}]_{m\cdot} &\sim\mathcal{D}irichlet_M(\underline{e}, \dots, \underline{e})\\
\boldsymbol{\pi}_0 &\sim\mathcal{D}irichlet_M(\underline{e}_0, \dots, \underline{e}_0).\label{eq:pi0prior}
\end{align}

The package \pkg{bsvars} offers two alternative models based on MSH. The
first is characterised by a stationary Markov process with no absorbing
state, and with a positive minimum number of regime occurrences
following \cite{dww2017}. In this model, the hyper-parameter
\(\underline{e} = \underline{e}_0\) is fixed to 1 by default and can be
modified by the user. The other model is a novel proposal by
\cite{SW2026} of a sparse representation that fixes the number of
regimes to an over-fitting value \(M=20\) (or specified by the user). In
this model, many of the regimes will have zero occurrences throughout
the sample, which allows the number of regimes with non-zero occurrences
to be estimated following the ideas by \cite{malsiner2016model}. Its
prior specification is complemented by a hierarchical prior for the
hyper-parameter \(\underline{e}\): \begin{align}
\underline{e} \sim \mathcal{IG}2\left(\underline{s}_e, \underline{\nu}_e\right). \label{eq:sparseprior}
\end{align} Due to its construction, the sparse MSH model excludes
regime-specific interpretation of parameters. Instead, the estimated
sequence of conditional variances, \(\sigma_{n.t}^2\), enjoys standard
interpretations. Furthermore, both MSH models provide identification
through heteroskedasticity following the ideas by \cite{LLM2010} and
\cite{LW2017}. The latter paper provides framework for verifying the
identification, which in the \pkg{bsvars} package is implemented by
verifying the homoskedasticity hypothesis represented by a restriction
setting the conditional variances to 1, \(\sigma_{n.m}^2 =  1\) for all
\(m\).

\subsection{Models with Non-Normal Shocks}

Identification of the structural shocks through non-normality is
implemented in the package using the mixture of normal component model
(MIX) following the proposal by \cite{Lanne2010} and the Student-t
distributed shocks as suggested by \cite{lanne2017}.

\subsubsection{Mixture of normal components}

In this model, the structural shocks follow a conditional \(N\)-variate
normal distribution given the state variable \(s_t=m\in\{1,\dots,M\}\);
\begin{align}
u_t\mid \mathbf{x}_t, s_t=m, \boldsymbol{\sigma}_m^2 \sim\mathcal{N}_N\left(\mathbf{0}_N, \text{diag}\left(\boldsymbol{\sigma}_m^2\right)\right),
\end{align} and where the states are predicted to occur in the next
period with probability \(\boldsymbol\pi_0\). The prior specification
for these models closely follows that for the MSH models, with the
Dirichlet prior for the regime probabilities, \(\boldsymbol\pi_0\), as
in \eqref{eq:pi0prior}, and that for the conditionals variances,
\(\boldsymbol{\sigma}_m^2\), as in \eqref{eq:sigmaMSprior}.

As long as the predictive state probabilities are constant in these
models the classification of the observations into the regimes is
performed using filtered and smoothed probabilities, or the posterior
realisations of the state allocations \(s_t\)
\citep[see][for a recent review of the methods]{song2021markov}.

The MIX model comes in two versions as well. The first is the finite
mixture model \citep[see e.g.][]{FruhwirthSchnatter2006} in which the
number of states, \(M\), is fixed and the unconditional state
probabilities, \(\boldsymbol\pi_0\), are strictly positive. The latter
condition requires non-zero regime occurrences over the sample, a
condition that is imposed in the package implementation. An alternative
specification is referred to as the sparse mixture model and is based on
the proposal by \cite{malsiner2016model}. In this model, the number of
the finite mixture components is set to be larger than the real number
of the components. Consequently, the number of components with non-zero
probability of occurrence is estimated, which is facilitated by allowing
the remaining components to have zero occurrences. This sparse structure
of normal components implemented thanks to the prior specified for the
hyper-parameter \(\underline{e}\) of Dirichlet distribution as in
equation \eqref{eq:sparseprior}.

The MIX models facilitate the identification through non-normality as
proposed by \cite{Lanne2010}. The hypothesis of normality for a shock,
contradicting its identification, is verified by checking whether
restriction \(\sigma_{n.m}^2 =  1\) holds for all \(m\).

\subsubsection{Student-t shocks}

The Bayesian implementation of the Student-t model follows closely that
by \cite{g93} and is implemented using an inverse gamma scale of normal
distribution. Therefore, the conditional normality of the structural
shocks from equation \eqref{eq:ss} is complemented by the marginal prior
distributions for variances \begin{align}
\sigma_{n.t}^2\mid \nu_n \sim\mathcal{IG}2(\nu_n - 2, \nu_n).
\end{align} Such a construction of the structural shock density results
in a marginal density for the \(n\)th shock being a zero-mean,
unit-variance Student-t distribution with equation-specific degrees of
freedom, \(\nu_n > 2\), \citep[see][]{Bauwens1999} \begin{align}
u_{n.t}\mid\mathbf{x}_t, \nu_n \sim t\left(0, 1, \nu_n\right).
\end{align} In this model, the only role of \(\sigma_{n.t}^2\) is to be
integrated out for the sake of specifying the demanded marginal density
for the structural shocks.

The prior distribution for the degrees of freedom parameter is set to
\begin{align}
p(\nu_n) = \frac{1}{(\nu_n - 1)^2}.
\end{align} This prior density is proper and setting it implies the
estimation of the degrees of freedom parameters. Its particular form is
further motivated by the fact that it facilitates identification through
non-normality verification for a shock by checking the restriction
\(\nu\rightarrow\infty\) \citep[see also][]{JENSEN20133}.

\subsection{Hypothesis Verification Using SDDRs}

The \pkg{bsvars} package includes unique procedures for the verification
of identification and hypotheses on autoregressive parameters. They are
based on posterior odds ratio computed using the SDDR by
\cite{Verdinelli1995}. Consider a general specification of a hypothesis
represented by sharp restrictions on the parameters of the model,
denoted by \(\mathcal{H}_0\), and its complement denoted by
\(\mathcal{H}_1\). The SDDR is specified by \begin{align}
\frac{\Pr[\mathcal{H}_0\mid data]}{\Pr[\mathcal{H}_1\mid data]} = SDDR = \frac{p(\mathcal{H}_0\mid data)}{p(\mathcal{H}_0)},\label{eq:sddr}
\end{align} where the LHS equality represents its interpretation as the
posterior odds ratio, whereas the RHS equality provides the equivalent
form that makes its computation straightforward. Consequently, SDDRs
report the ratio of the posterior probability of the restriction to the
posterior probability of the unrestricted model. Therefore, a value of
the SDDR greater than one provides evidence in favour of the restriction
\(\mathcal{H}_0\), whereas its value less than one provides evidence
against this hypothesis, or in favour of \(\mathcal{H}_1\).

The SDDR computation requires the estimation of the unrestricted model
under \(\mathcal{H}_1\) and the computation of the ratio of the marginal
posterior ordinate to the marginal prior ordinate both evaluated at the
restriction \(\mathcal{H}_0\). The specification of the models included
in the \pkg{bsvars} package facilitate fast estimation of both ordinates
using the estimator proposed by \cite{gs90}.

\subsubsection{Identification verification}

In order to verify the identification of structural shocks through
time-varying volatility (or non-normality) one needs to verify the
hypothesis of homoskedasticity (normality) of individual shocks, which
allows them to make probabilistic statements regarding partial or global
identification \citep[see][]{LW2017,LSUW2024,lanne2017}. The model is
globally identified iff no more than one structural shock is
homoskedastic (normal). An individual shock is identified iff it is
heteroskedastic (non-normal) or if it is the only homoskedastic (normal)
shock in the system.

According to \cite{LSUW2024} the hypothesis of homoskedasticity of the
\(n\)th shock in the non-centred SV model is represented by the
restriction \begin{align}
\mathcal{H}_0:\quad \omega_n = 0
\end{align} whereas in the MSH models, according to \cite{SW2026}, it
given by \begin{align}
\mathcal{H}_0:\quad \sigma_{n.1}^2 = \dots \sigma_{n.M}^2 = 1.\label{eq:homoMSH}
\end{align}

The verification of normality of the \(n\)th shock in the MIX models is
performed using the restriction in equation \eqref{eq:homoMSH}, whereas
in the Student-t model it is given by \begin{align}
\mathcal{H}_0:\quad \nu_n \rightarrow\infty,
\end{align} as in the limit, the Student-t distribution becomes normal.
The model specification makes Bayesian verification of this uncommon
restriction straightforward.

\subsubsection{Verifying autoregressive specification}

Finally, the package makes it possible to verify restrictions on the
autoregressive parameters in the form of \begin{align}
\mathcal{H}_0:\quad \mathbf{S}\text{vec}(\mathbf{A}) = \mathbf{r},
\end{align} where \(\text{vec}(\mathbf{A})\) is a vectorised matrix
\(\mathbf{A}\), \(\mathbf{S}\) is an \(r\times N(Np+D)\) selection
matrix picking \(r\) elements of \(\mathbf{A}\) to be restricted to the
values in the \(r\)-vector \(\mathbf{r}\). Note that the specification
of the hierarchical prior leading to the estimated level of
autoregressive shrinkage makes the verification of such restrictions
less depending on arbitrary choices.

\subsection{Posterior Samplers and Computational Details}\label{sec:posterior}

In this section, we explain \pkg{bsvars} package's implementation of
fast and efficient estimation algorithms obtained thanks to the
application of appropriate model specification and frontier econometric
techniques best described in \cite{LSUW2024}.

The objective for choosing the model equations and the prior
distributions was to make the estimation using Gibbs sampler technique
\citep[see e.g.][]{CasellaGeorge1992} and well-specified
easy-to-sample-from full conditional posterior distributions. Therefore,
the package relies on the reduced form equation \eqref{eq:rf} for the
VAR model. This choice is fairly uncommon in the SVAR literature but it
simplifies the estimation of the autoregressive parameters
\(\mathbf{A}\) directly in the form as they are used for the
computations of impulse responses or forecast error variance
decomposition. This choice combined with the prior in \eqref{eq:priorA}
and specification of the structural form equation \eqref{eq:sf}
facilitates the application of the row-by-row sampler by
\cite{chankoopyu2024}. As shown by \cite{CARRIERO2022}, relative to the
joint estimation of the matrix \(\mathbf{A}\) in one step, a usual
practice in reduced form VARs, the row-by-row estimation in SVARs can
reduce computational complexity of Bayesian estimation from
\(\mathcal{O}(N^6)\) to \(\mathcal{O}(N^4)\).

The estimation of the structural form equation \eqref{eq:sf} is
implemented following the quickly converging, efficient, and providing
excellent mixing sampling algorithm by \cite{WaggonerZha2003}. It offers
a flexible framework for setting exclusion restrictions and is adapted
to the SVARs identified through heteroskedasticity and non-normality.
The unique formulation of this equation is particularly convenient for
complex heteroskedastic models facilitating the row-by-row estimation of
matrix \(\mathbf{A}\) and Gibbs sampler for the heteroskedastic process.

The estimation of the Stochastic Volatility models is particularly
requiring due to the \(N\) independent \(T\)-valued latent volatility
processes estimation that it involves. The implementation of crucial
techniques is particularly important here. The sampling algorithms use
the 10-component auxiliary mixture technique by \cite{Omori2007} that
facilitates the estimation of the log-volatility using the simulation
smoother by \cite{mccausland2011simulation} for conditionally Gaussian
linear state-space models greatly speeding up the computations
\citep[see][for the computational times comparison for various estimation algorithms]{twss_2021}.
Application of appropriate numerical techniques reduces the complexity
from \(\mathcal{O}(T^3)\) to \(\mathcal{O}(T)\).

Additionally, our specification facilitates the algorithms to estimate
heteroskedastic process if the signal from the data is strong, but it
also allows them to heavily shrink the posterior towards
homoskedasticity, as in equation \eqref{eq:homosk}, otherwise. The
package implements the adaptation of the ancillarity-sufficiency
interweaving strategy that is shown by \cite{Kastner2014} to improve the
efficiency of the sampler when heteroskedasticity is uncertain. Our
implementation of the sampling algorithm closely follows the algorithms
from package \pkg{stochvol} with adaptations necessary for the SVAR
modelling.

The estimation of the Markov switching and mixture models benefits
mainly from the implementation of the forward-filtering
backward-sampling estimation algorithm for the Markov process \(s_t\) by
\cite{Chib1996} in \proglang{C++}. However, an additional step of
choosing the parameterisation of the conditional variances as in
equation \eqref{eq:sigmaMSprior}, requiring sampling from a new
distribution, assures excellent mixing and sampling efficiency
improvements relative to alternative ways of standardising these
parameters.

All of the estimation routines for the Markov chain Monte Carlo
estimation of the models and those for low level processing of the rich
estimation output are implemented based using compiled code in
\pkg{C++}. This task is facilitated by the \pkg{Rcpp} package by
\cite{eddelbuettel2011rcpp} and \cite{eddelbuettel_seamless_2013}. The
\pkg{bsvars} package relies heavily on linear algebra and pseudo-random
number generators. The former is implemented using the package
\pkg{RcppArmadillo} by \cite{eddelbuettel_rcpparmadillo2014} that is a
collection of headers linking to the \proglang{C++} library
\pkg{armadillo} by \cite{sanderson2016armadillo}, as well as on several
utility functions for operations on tri-diagonal matrices from package
\pkg{stochvol} by \cite{factorstochvol}. The latter refers to the
algorithms from the standard normal distribution using package
\pkg{RcppArmadillo}, truncated normal distribution \pkg{RcppTN} by
\cite{RcppTN} implementing the efficient sampler by \cite{Robert1995},
and generalised inverse Gaussian distribution using package \pkg{GIGrvg}
by \cite{GIGrvg} implementing the sampler by
\cite{hormann2014generating}. All of these developments make the
algorithms computationally fast. Still, Bayesian estimation of
multivariate dynamic structural models is a requiring task that might
take a little while. To give users a better idea of the remaining time
the package displays a progress bar implemented using the package
\pkg{RcppProgress} by \cite{RcppProgress}. Finally, the rich structure
of the model specification including the prior distributions,
identification pattern, and starting values, as well as the rich outputs
from the estimation algorithms are organised using dedicated classes
within the \pkg{R6} package by \cite{R6} functionality.

\section{Workflows for SVAR analysis}

The \pkg{bsvars} package allows the users to design their workflows in
several ways. The three main stages include model specification,
estimation, and post-estimation analysis.

\subsection{The Basic Workflow (also with a Pipe)}

The basics of the workflow are presented for the case of a simple
homoskedastic SVAR model. Begin by uploading the package and a sample
data matrix for the analysis of the US fiscal policy, and setting the
seed for the sake of reproducibility:

\begin{CodeChunk}
\begin{CodeInput}
R> library(bsvars)
R> data(us_fiscal_lsuw)
R> set.seed(1)
\end{CodeInput}
\end{CodeChunk}

Specify the SVAR model with a lower triangular structural matrix and one
autoregressive lag both being the default settings by executing:

\begin{CodeChunk}
\begin{CodeInput}
R> spec = specify_bsvar$new(us_fiscal_lsuw)
\end{CodeInput}
\begin{CodeOutput}
The identification is set to the default option of lower-triangular structural matrix.
\end{CodeOutput}
\end{CodeChunk}

The object \texttt{spec} includes a number of objects that define the
model, such as the data matrices, fixed hyper-parameters of the prior
distribution, identification pattern, and starting values. The
estimation of the model using the Monte Carlo Markov Chain (MCMC)
methods is performed in two steps. First, the object \texttt{spec} with
model specification is provided to the \texttt{estimate()} function to
perform \texttt{1000} iterations of the Gibbs sampler. This burn-in
stage is run for the algorithm to achieve convergence to the stationary
posterior distribution. In the second step, the object \texttt{burn} is
provided to the \texttt{estimate()} function to perform \texttt{10000}
iterations of the Gibbs sampler and save them in the object
\texttt{post}:

\begin{CodeChunk}
\begin{CodeInput}
R> burn = estimate(spec, 1000)
R> post = estimate(burn, 10000)
\end{CodeInput}
\end{CodeChunk}

In the latter case, the function \texttt{estimate()} extracts the last
draw of the parameters from the previous run, uses it as starting values
for the final run of the estimation algorithm, and continues the Markov
chain providing sample from the target posterior distribution. This way
of designing the \texttt{estimate()} function allows the user to run the
estimation in several steps, which is particularly useful for the models
requiring longer runs of the sampling algorithm to achieve convergence.
The user can discard the draws from the previous step and continue the
estimation in as many subsequent \texttt{estimate()} function runs as
required.

The posterior output is normalised with respect to the signs of the
structural matrix following the procedure by \cite{WaggonerZha2003norm}
in an automated way by running function \texttt{normalise()} within the
execution of the function \texttt{estimate()}.

Subsequently, use the estimation output to forecast, say, four periods
ahead:

\begin{CodeChunk}
\begin{CodeInput}
R> fore = forecast(post, 4); summary(fore)
\end{CodeInput}
\end{CodeChunk}

compute and plot impulse responses:

\begin{CodeChunk}
\begin{CodeInput}
R> irfs = compute_impulse_responses(post, 4); plot(irfs)
\end{CodeInput}
\end{CodeChunk}

verify whether the third variable Granger causes the first:

\begin{CodeChunk}
\begin{CodeInput}
R> H0   = matrix(NA, 3, 4); H0[1,3] = 0
R> sddr = verify_autoregression(post, H0)
R> summary(sddr)
\end{CodeInput}
\end{CodeChunk}

or perform any other post-estimation analysis.

The same workflow can be coded using the
\texttt{\textbar{}\textgreater{}} pipe. We propose that the user binds
the model specification with the estimation:

\begin{CodeChunk}
\begin{CodeInput}
R> set.seed(1)
R> us_fiscal_lsuw |> 
+   specify_bsvar$new() |> 
+   estimate(S = 1000) |> 
+   estimate(S = 10000) -> post
\end{CodeInput}
\end{CodeChunk}

to obtain exactly the same draws from the posterior distribution as in
the previous workflow collected in object \texttt{post}. The
post-estimation analysis can then be executed by:

\begin{CodeChunk}
\begin{CodeInput}
R> post |> forecast(horizon = 4) |> summary()
R> post |> compute_impulse_responses(horizon = 4) |> plot()
R> post |> verify_autoregression(hypothesis = H0) |> summary()
\end{CodeInput}
\end{CodeChunk}

User preferences, convenience, and specific project requirements should
decide on the implementation of the procedure in a script.

\subsection{Customizing The Workflow}

\begin{figure}[ht!]
\begin{center}
\begin{tikzpicture}

\node[draw,
        minimum width=4cm,
        minimum height=1cm] 
    (specify) 
    {\begin{tabular}{l}
    \textit{Specify a model}\\[1ex]
    \verb|specify_bsvar| \\
    \verb|specify_bsvar_mix|\\
    \verb|specify_bsvar_msh|\\
    \verb|specify_bsvar_sv|\\
    \verb|specify_bsvar_t|
    \end{tabular}};

\node[draw,
      below=of specify,
        minimum width=4cm,
        minimum height=1cm] 
    (estimate) 
    {\begin{tabular}{l}
    \textit{Estimate a model}\\[1ex]
    \verb|estimate|
    \end{tabular}};

\node[draw,
      right=of estimate,
        minimum width=5cm,
        minimum height=1cm] 
    (continue) 
    {\begin{tabular}{l}
    \textit{Continue the MCMC}\\[1ex]
    \verb|estimate|
    \end{tabular}};

\node[draw,
      left=of estimate,
        minimum width=4cm,
        minimum height=1cm] 
    (normalise) 
    {\begin{tabular}{l}
    \textit{Normalise posterior}\\[1ex]
    \verb|normalise|
    \end{tabular}};

\node[draw,
      below right=of estimate,
        minimum width=4cm,
        minimum height=1cm] 
    (forecast) 
    {\begin{tabular}{l}
    \textit{Forecast}\\[1ex]
    \verb|forecast|
    \end{tabular}};

\node[draw,
      below=of forecast,
        minimum width=4cm,
        minimum height=1cm] 
    (verify) 
    {\begin{tabular}{l}
    \textit{Verify hypotheses}\\[1ex]
    \verb|verify_autoregression|\\
    \verb|verify_identification|
    \end{tabular}};

\node[draw,
      below=of verify,
        minimum width=4cm,
        minimum height=1cm] 
    (compute) 
    {\begin{tabular}{l}
    \textit{Compute interpretable quantities}\\[1ex]
    \verb|compute_conditional_sd|\\
    \verb|compute_fitted_values|\\
    \verb|compute_historical_decompositions|\\
    \verb|compute_impulse_responses|\\
    \verb|compute_regime_probabilities|\\
    \verb|compute_structural_shocks|\\
    \verb|compute_variance_decompositions|
    \end{tabular}};

\draw[-latex] (specify) edge (estimate);
\draw[-latex] (estimate) |- (forecast);
\draw[-latex] (estimate) |- (compute);
\draw[-latex] (estimate) |- (verify);

\draw[arrows={Stealth[scale=0.8,inset=0pt]-}] (estimate) to [bend left=45] (continue);
\draw[arrows={-Stealth[scale=0.8,inset=0pt]}] (estimate) to [bend right=45] (continue);

\draw[arrows={-Stealth[scale=0.8,inset=0pt]}] (estimate) to [bend left=45] (normalise);
\draw[arrows={Stealth[scale=0.8,inset=0pt]-}] (estimate) to [bend right=45] (normalise);

\end{tikzpicture}
\end{center}
\caption{Workflow for SVAR analysis}
\label{fig:workflow}
\end{figure}
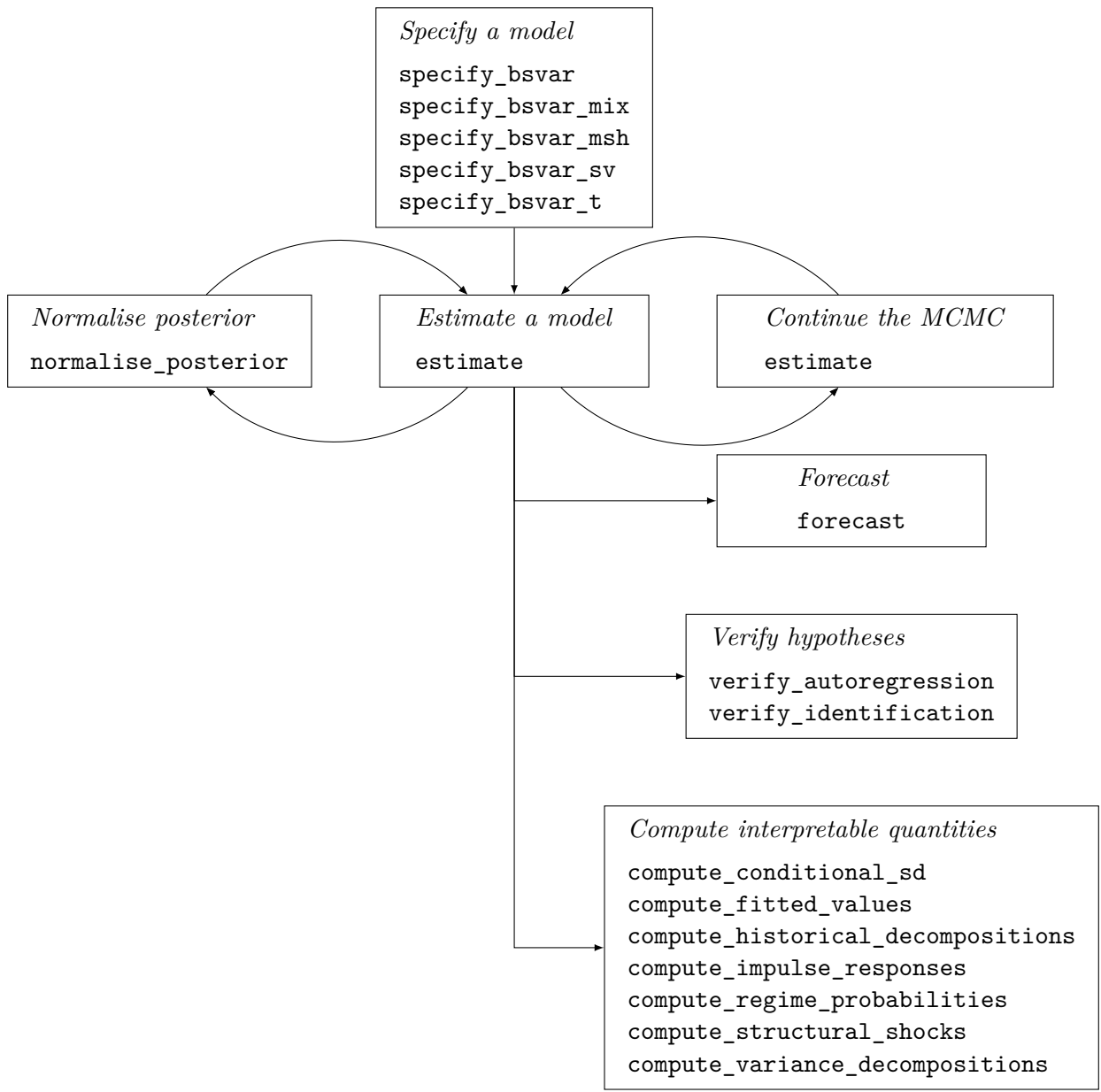

The package offers a range of models, post-estimation methods, and a set
of functions for each of the stages of analysis. The full extent of
workflow customization is presented in Figure \ref{fig:workflow}.

The top of the Figure presents functions to specify the model. Here
users can choose a homoskedastic model, one with normal mixture (MIX),
Markov-switching heteroskedasticity (MSH), Stochastic Volatility (SV),
or t-distributed errors. Then the model can be estimated using the
\texttt{estimate()} function repeatedly as long as necessary to obtain
\(S\) draws from the posterior distribution in the final run. It is
recommended then that the user verifies the draws of the \(N\times N\)
structural matrix contained in an \(N\times N\times S\) array
\texttt{post\$posterior\$B} for uni-modality. The lack of thereof may
indicate that the automated manner of normalising the posterior draws
using function \texttt{normalise()} was not successful and more work
needs to be done here. The default calibration of this algorithm is
found to be sufficient for models with exclusion restrictions though.

Given the posterior output, the user can proceed to forecasting,
hypotheses verification, or analysis of interpretable quantities. No
particular ordering of the operations is recommended and users should
implement their competence in the subject matter and consider their
project's requirements here. Forecasting is performed using the
\texttt{forecast()} function and includes the possibility of generating
conditional predictions given future projections of some of the
variables provided in argument \texttt{conditional\_forecast}. The user
can verify hypotheses about the model parameters using the
\texttt{verify\_autoregression()} and \texttt{verify\_identification()}
functions implementing the computation of SDDRs. Finally, the user can
compute interpretable quantities such as structural shocks' conditional
standard deviations, fitted values, historical decompositions, impulse
responses, MIX and MSH models' regime probabilities, structural shocks,
and forecast error variance decompositions using an appropriate method
listed in the last block in Figure \ref{fig:workflow}. The package
documentation provides all the necessary details on the arguments and
outputs of the functions, which enables further customization of the
workflow.

\subsection{Generics, Methods, and Functions}

\begin{table}[ht!]
    \caption{Summary of package's generics, methods, and functions}
    
    \label{tab:gen}
    \begin{center}
    \begin{tabular}{lllll}
    \toprule
generic or function& first argument class & output class & \rotatebox[origin=c]{-90}{\texttt{plot}} & \rotatebox[origin=c]{-90}{\texttt{summary}}\\[1ex]
  \midrule
\multicolumn{5}{c}{\textit{Specify a model}}\\[1ex]
\verb|specify_bsvar|&\verb|matrix|&\verb|BSVAR|, \verb|R6|&&\\
\verb|specify_bsvar_mix|&\verb|matrix|&\verb|BSVARMIX|, \verb|R6|&&\\
\verb|specify_bsvar_msh|&\verb|matrix|&\verb|BSVARMSH|, \verb|R6|&&\\
\verb|specify_bsvar_sv|&\verb|matrix|&\verb|BSVARSV|, \verb|R6|&&\\
\verb|specify_bsvar_t|&\verb|matrix|&\verb|BSVART|, \verb|R6|&&\\[1ex]
  \midrule
\multicolumn{5}{c}{\textit{Estimate a model}}\\[1ex]
\multirow{2}{*}{\texttt{estimate}} &(model classes)&(posterior classes)&&\checkmark\\
&(posterior classes)&(posterior classes)&&\checkmark\\[1ex]
  \midrule
\multicolumn{5}{c}{\textit{Normalise posterior output (if necessary)}}\\[1ex]
\verb|normalise| &(posterior classes)&(posterior classes)&&\\[1ex]
  \midrule
\multicolumn{5}{c}{\textit{Forecast}}\\[1ex]
\verb|forecast| &(posterior classes)&\verb|Forecasts|&\checkmark&\checkmark\\[1ex]
  \midrule
\multicolumn{5}{c}{\textit{Compute interpretable quantities}}\\[1ex]
\verb|compute_conditional_sd| &(posterior classes)&\verb|PosteriorSigma|&\checkmark&\checkmark\\
\verb|compute_fitted_values| &(posterior classes)&\verb|PosteriorFitted|&\checkmark&\checkmark\\
\verb|compute_historical_decompositions| &(posterior classes)&\verb|PosteriorHD|&\checkmark&\checkmark\\
\verb|compute_impulse_responses| &(posterior classes)&\verb|PosteriorIR|&\checkmark&\checkmark\\
\verb|compute_regime_probabilities| &(posterior classes)&\verb|PosteriorRegimePr|&\checkmark&\checkmark\\
\verb|compute_structural_shocks| &(posterior classes)&\verb|PosteriorShocks|&\checkmark&\checkmark\\
\verb|compute_variance_decompositions| &(posterior classes)&\verb|PosteriorFEVD|&\checkmark&\checkmark\\[1ex]
  \midrule
\multicolumn{5}{c}{\textit{Verify hypotheses}}\\[1ex]
\verb|verify_autoregression| &(posterior classes)&\verb|SDDRautoregression|&&\checkmark\\
\verb|verify_identification| &(posterior classes)&(sddr classes)&&\checkmark\\[1ex]
  \midrule
\multicolumn{5}{c}{\textit{Classes explanation}}\\[1ex]
  \multicolumn{5}{p{\linewidth}}{(model classes) include \texttt{BSVAR}, \texttt{BSVARMIX}, \texttt{BSVARMSH}, \texttt{BSVARSV}, \texttt{BSVART}} \\
  \multicolumn{5}{p{\linewidth}}{(posterior classes) include \texttt{PosteriorBSVAR}, \texttt{PosteriorBSVARMIX}, \texttt{PosteriorBSVARMSH}, \texttt{PosteriorBSVARSV}, \texttt{PosteriorBSVART}} \\
  \multicolumn{5}{p{\linewidth}}{(sddr classes) include \texttt{SDDRidSV}, \texttt{SDDRidMIX}, \texttt{SDDRidMSH}, \texttt{SDDRidT}} \\[1ex]
    \bottomrule
    \end{tabular}
    \end{center}
Note: The last two columns indicate availability a dedicated \texttt{plot} or \texttt{summary} method.
\end{table}

The workflows described in this section are possible thanks to a
deliberate design of the package's generics, methods, and functions
listed in Table \ref{tab:gen}. The functions specifying the model are
created using the \pkg{R6} classes to facilitate the management of this
complicated object. All the specification details can be adjusted by the
user by modifying the elements of the specification object, such as the
\texttt{spec} object, that is of a class indicating the specified model,
e.g.~\texttt{BSVAR}, \texttt{BSVARSV}, \ldots{} . These objects inherit
properties of the \texttt{list} class as well.

All other functions listed in Table \ref{tab:gen} define generics that
find their particular methods depending on the class of their first
argument. Such a construction assures that appropriate estimation
algorithm is applied to the model specified by the user. This applies,
to all the subsequent stages of the workflow, such as forecasting,
hypotheses verification, and computation of interpretable quantities
that are performed using the exact algorithms required by the particular
specification.

This can be illustrated for the workflow described in the current
section. Executing the specification function
\texttt{specify\_bsvar\$new()} creates an object of class
\texttt{BSVAR}. Therefore, in the second step the \texttt{estimate()}
generic implements method for homoskedastic SVAR model coded in
\texttt{estimate.BSVAR()} creating object of class
\texttt{PosteriorBSVAR}, which is followed by execution of the method
\texttt{estimate.PosteriorBSVAR()} that continues the model estimation
using MCMC methods. Consequently, the forecasting is performed using
method \texttt{forecast.PosteriorBSVAR()}, impulse responses are
computed for this model using method
\texttt{compute\_impulse\_responses.PosteriorBSVAR()}, and the SDDR is
estimated using the algorithm dedicated to this particular model using
method \texttt{verify\_autoregression.PosteriorBSVAR()}. A dedicated set
of \texttt{summary()} and \texttt{plot()} methods greatly simplifies the
user experience and their workflows.

Finally, generics are designed to be used in other packages with the
implementation in the \proglang{R} packages \pkg{bsvarSIGNs},
\pkg{bvars}, and \pkg{bpvars}.

\section{Conclusion}

The \pkg{bsvars} package offers fast and efficient algorithms for
Bayesian estimation of Structural VARs with a range of specifications
for the volatility or distributions of the structural shocks. Thanks to
the application of the frontier econometric and numerical techniques the
package makes the estimation of multivariate dynamic structural models
feasible even for a larger number of variables, complex identification
strategies, and non-linear specifications. Its strong reliance on
algorithms written in \proglang{C++} makes it possible to benefit from
the best of the two worlds: the convenience of data analysis using
\proglang{R} and the computational speed using pre-compiled code written
in \proglang{C++}. Finally, the package provides essential generics for
applied analyses that facilitate developing coherent workflows for the
dependent packages, such as \pkg{bsvarSIGNs}, \pkg{bvars}, and
\pkg{bpvars}, package greatly extending the set of models available to
the users.

\bibliography{bsvars.bib}

\end{document}